\documentstyle[12pt,a4]{article}
\input epsf
\global\arraycolsep=2pt

\begin{document}

\begin{titlepage}

\vbox{}
\vspace{3.0cm}

\begin{center}
\Large\bf Test of the Running of $\alpha_s$ in $\tau$ Decays
\end{center}

\vspace{1.0cm}

\begin{center}
Maria Girone\\
{\sl Dipartimento di Fisica, INFN Sezione di Bari, 70126 Bari,
Italy}\\
\vspace{0.5cm}
and\\
\vspace{0.5cm}
Matthias Neubert\\
{\sl Theory Division, CERN, CH-1211 Geneva 23, Switzerland}
\end{center}

\vspace{1.2cm}

\begin{abstract}
The $\tau$ decay rate into hadrons of invariant mass smaller than
$\sqrt{s_0}\gg\Lambda_{\rm QCD}$ can be calculated in QCD assuming
global quark--hadron duality. It is shown that this assumption holds
for $s_0>0.7$~GeV$^2$. From measurements of the hadronic mass
distribution, the running coupling constant $\alpha_s(s_0)$ is
extracted in the range 0.7~GeV$^2<s_0<m_\tau^2$. At $s_0=m_\tau^2$,
the result is $\alpha_s(m_\tau^2)=0.329\pm 0.030$. The running of
$\alpha_s$ is in good agreement with the QCD prediction.
\end{abstract}

\vspace{1.5cm}
\centerline{(Phys.\ Rev.\ Lett.\ {\bf 76} (1996) 3061)}

\end{titlepage}

The scale dependence of coupling constants is one of the key features
of renormalizable quantum field theories. In QCD, the effective
coupling constant $\alpha_s(Q^2)$ is predicted to decrease with the
momentum transfer $Q^2$, a property referred to as asymptotic freedom
\cite{Gros}. This prediction has been tested by comparing data
obtained from experiments operating at different energies
\cite{Webb}; it has also been studied in single high-energy
experiments at $e p$ and $p\bar p$ colliders, where a large range in
$Q^2$ can be probed simultaneously \cite{Giel}. Here we propose a
test of the scale dependence of $\alpha_s(Q^2)$ in the low-energy
region 0.7~GeV$^2<Q^2<m_\tau^2$. Our method is based on integrals of
the invariant mass distribution in hadronic $\tau$ decays. It
provides a unique opportunity to test one of the most important
predictions of QCD in a single experiment and at low energies, where
the effect of the running of $\alpha_s$ is most pronounced.

We shall consider the $\tau$ decay rate into hadrons of invariant
mass squared smaller than $s_0$, normalized to the leptonic
decay rate:
\begin{equation}
   R_\tau(s_0) = {\Gamma(\tau\to\nu_\tau + \mbox{hadrons};\,
   s_{\rm had} < s_0)\over\Gamma(\tau\to\nu_\tau\,e\,\bar\nu_e)}
   = \int\limits_0^{\displaystyle s_0}\!{\rm d}s\,
   {{\rm d}R_\tau(s)\over{\rm d}s} \,,
\end{equation}
where ${\rm d}R_\tau/{\rm d}s$ is the inclusive hadronic spectrum. As
long as $s_0\gg\Lambda_{\rm QCD}^2$, the quantity $R_\tau(s_0)$ can
be calculated in QCD using the Operator Product Expansion (OPE)
\cite{SVZ,Rtau1}. Applying the OPE in the physical region assumes
global quark--hadron duality, i.e.\ that decay rates admit a QCD
description after a ``smearing'' over a sufficiently wide energy
interval has been performed \cite{PQW}, which in the present case is
provided by the integration over the range $0<s<s_0$. The question of
how accurate this assumption is and for what values of $s_0$ it
applies is a phenomenological one; it cannot be answered yet from
theoretical grounds. Below, we shall investigate this question,
comparing data with theoretical predictions based on the duality
assumption. A similar test of duality has been performed in
Ref.~\cite{epem}, using data on the $e^+ e^-\to\mbox{hadrons}$ cross
section.

The $\tau$ decay rate into hadrons can be written in terms of moments
${\cal M}_k^{(J)}$ of the absorptive part of current--current
correlation functions of angular momentum $J$ \cite{LP2,MN}. The
quantity $R_\tau(s_0)$ is given by
\begin{eqnarray}\label{Rs0}
   {1\over 3 S_{\rm EW}}\,R_\tau(s_0)
   &=& {2 s_0\over m_\tau^2}\,{\cal M}_0^{(1)}(s_0)
    - 2\,\bigg( {s_0\over m_\tau^2}\bigg)^3\,{\cal M}_2^{(1)}(s_0)
    + \bigg( {s_0\over m_\tau^2}\bigg)^4\,{\cal M}_3^{(1)}(s_0)
    \nonumber\\
   &+& {2 s_0\over m_\tau^2}\,{\cal M}_0^{(0)}(s_0)
    - 2\,\bigg( {s_0\over m_\tau^2}\bigg)^2\,{\cal M}_1^{(0)}(s_0)
    + {2\over 3}\,\bigg( {s_0\over m_\tau^2}\bigg)^3\,
    {\cal M}_2^{(0)}(s_0) \,,
\end{eqnarray}
where $S_{\rm EW}\simeq 1.0194$ accounts for electroweak radiative
corrections \cite{Sirl}. The moments can be written as contour
integrals along a circle of radius $s_0$ in the complex plane. Since
the only large mass scale in these integrals is $s_0$, the OPE
provides an expansion in powers of $1/s_0$:
\begin{equation}\label{OPE}
   {\cal M}_k^{(J)}(s_0)
   = {\cal M}_k^{(1)}[\alpha_s(s_0)]_{\rm pert}\,\delta_{J=1}
   + \sum_{n=1}^\infty c_n^{(J)}[\alpha_s(s_0)]\,
   {\langle O_{2n}\rangle\over s_0^n} \,.
\end{equation}
The leading term is given by perturbation theory alone. Terms
suppressed by powers of $1/s_0$ consist of perturbative coefficients
$c_n^{(J)}$ multiplying dimensionful parameters $\langle
O_{2n}\rangle$, such as quark masses or vacuum condensates
\cite{SVZ}. This is how nonperturbative effects are incorporated.
There is no leading term for the moments with $J=0$, which vanish in
the chiral limit and are thus proportional to powers of the light
quark masses. For the moments with $J=1$, the perturbative
contribution is
\begin{equation}\label{Mkpert}
   {\cal M}_k^{(1)}[\alpha_s(s_0)]_{\rm pert}
   = 1 + \sum_{n=1}^\infty d_n^{(k)}\,
   \bigg( {\alpha_s(s_0)\over\pi} \bigg)^n \,,
\end{equation}
where $\alpha_s(s_0)$ is defined in the $\overline{\mbox{\sc ms}}$
renormalization scheme, $d_1^{(k)}=1$, and the next three
coefficients are given by \cite{LP2,MN}
\begin{eqnarray}\label{dk}
   d_2^{(k)} &=& 1.63982 + {9\over 4(k+1)} \,, \nonumber\\
   d_3^{(k)} &=& -10.2839 + {11.3792\over k+1} + {81\over 8(k+1)^2}
    \,, \\
   d_4^{(k)} &=& K_4 - 155.955 - {46.238\over k+1}
    + {94.810\over(k+1)^2} + {68.344\over(k+1)^3} \,. \nonumber
\end{eqnarray}
The coefficient $K_4$ appears in the perturbative expansion of the
Adler function and is not known exactly. An estimate using the
principle of minimal sensitivity and the effective charge approach
\cite{ECH} gives $K_4\simeq 27.5$ \cite{Kata}. We shall use this
result in our analysis. The error due to the truncation of the
perturbation series in (\ref{Mkpert}) is of the order of the last
term included. It can also be estimated by summing a subset of
corrections to all orders in perturbation theory. Such a class of
corrections is provided by the renormalon chains \cite{tHof}, which
are the terms of order $\beta_0^{n-1}\alpha_s^n$, where $\beta_0$ is
the first coefficient of the $\beta$-function. For the case of the
moments, the resummation of these terms has been discussed in
Refs.~\cite{MN,BBB}. Below, we shall take fixed-order perturbation
theory as the nominal scheme and use the resummation of renormalon
chains to estimate the perturbative uncertainty.

The nonperturbative corrections in the OPE are proportional to the
light quark masses or to vacuum condensates \cite{SVZ}. We quote the
power corrections for the sum of the moments contributing to
$R_\tau(s_0)$ in (\ref{Rs0}). The terms relevant to the numerical
analysis are
\begin{eqnarray}\label{Rpower}
   {1\over 3 S_{\rm EW}}\,R_\tau(s_0)|_{\rm power}
   &=& - 6\,|\,V_{us}|^2\,{m_s^2(s_0)\over m_\tau^2}\,\bigg[
    1 + {s_0\over m_\tau^2}- \bigg( {s_0\over m_\tau^2}\bigg)^2
    + {1\over 3}\,\bigg( {s_0\over m_\tau^2}\bigg)^3 \bigg]
    \nonumber\\
   &&\mbox{}+ {16\pi^2\over m_\tau^4}\,\Big[
    \langle m_u\bar\psi_u\psi_u\rangle + |\,V_{ud}|^2\,
    \langle m_d\bar\psi_d\psi_d\rangle + |\,V_{us}|^2\,
    \langle m_s\bar\psi_s\psi_s\rangle \Big] \nonumber\\
   &&\mbox{}- {512\pi^3\over 27}\,
    {\rho\alpha_s\langle\bar\psi\psi\rangle^2\over m_\tau^6}
    + \dots \,,
\end{eqnarray}
where $m_s(s_0)$ is the running strange-quark mass, $\langle
m_q\bar\psi_q\psi_q\rangle$ are the quark condensates, and
$\rho\alpha_s \langle\bar\psi\psi\rangle^2$ denotes the four-quark
condensate. More detailed expressions, which are used in our
analysis, can be found elsewhere \cite{Rtau1,LP2,MN}. At tree level,
the powers of $1/s_0$ appearing in the OPE of the moments in
(\ref{OPE}) conspire with the powers of $s_0/m_\tau^2$, which
multiply the moments in (\ref{Rs0}), so that the nonperturbative
corrections to $R_\tau(s_0)$ are suppressed by powers of
$1/m_\tau^2$. This is no longer the case if radiative corrections to
the coefficients of the vacuum condensates are taken into account,
but the corresponding effects are very small. As a consequence, the
power corrections to $R_\tau(s_0)$ remain small down to rather low
values of $s_0$; using standard values of the QCD parameters (which
we take from Ref.~\protect\cite{MN}) we find $-(1.4\pm 0.5)\%$ for
the right-hand side of (\ref{Rpower}) at $s_0=m_\tau^2$, and
$-(1.5\pm 0.5)\%$ at $s_0=1$~GeV$^2$. This observation, together with
the fact that the perturbative contributions are known to high order,
guarantees a good convergence of the OPE down to low energy scales.

\begin{figure}
   \epsfysize=9cm
   \centerline{\epsffile{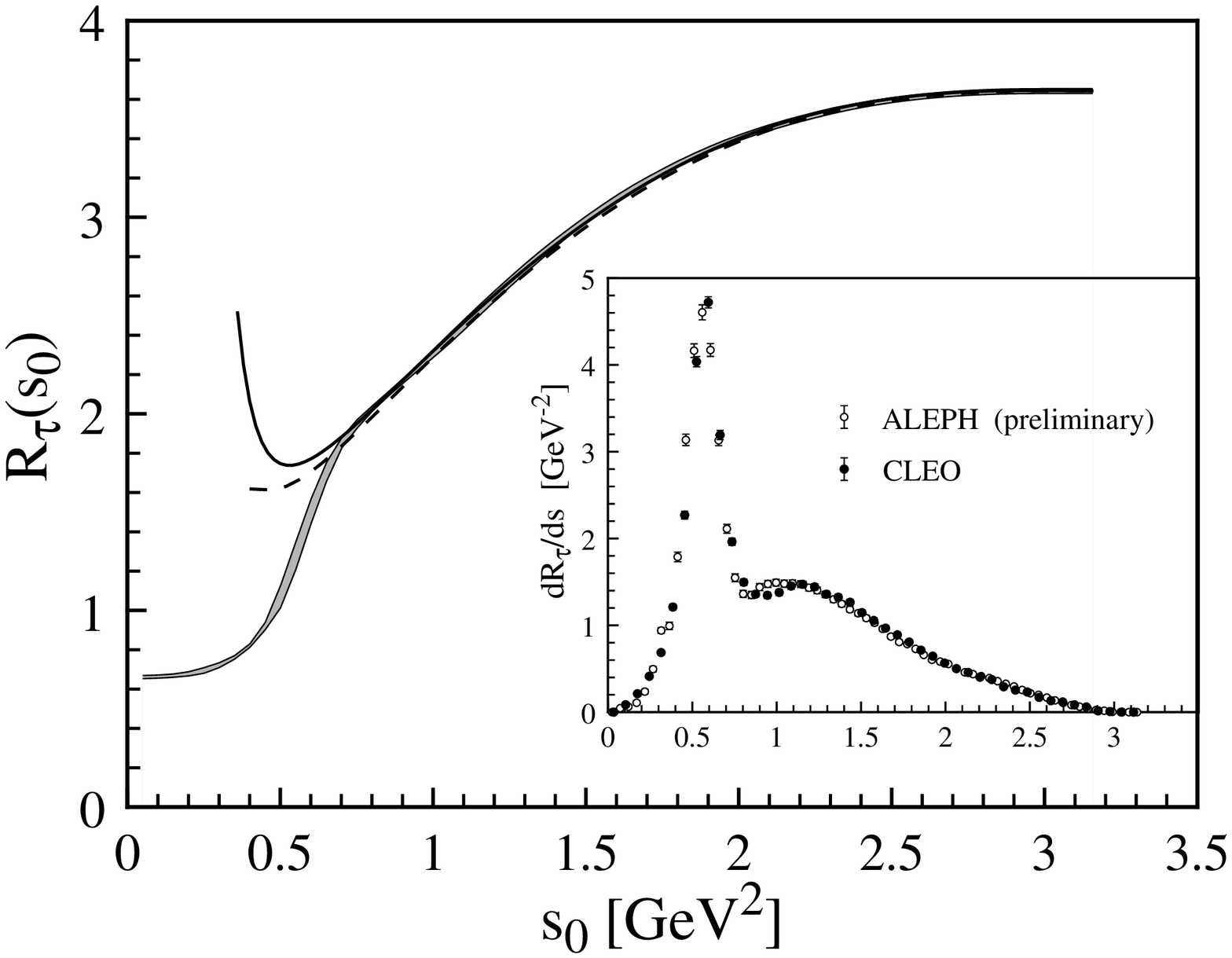}}
   \centerline{\parbox{14cm}{\caption{\label{fig:data}
The quantity $R_\tau(s_0)$ extracted from the data on the hadronic
mass distribution ${\rm d}R_\tau/{\rm d}s$ reported by the CLEO and
ALEPH Collaborations \protect\cite{CLEO,ALEPH} (small figure). The
experimental result is represented as a band. The curves show the
theoretical predictions (see text).
}}}
\end{figure}

To extract the quantity $R_\tau(s_0)$, we use the spectra of the
hadronic mass distribution reported by the CLEO and ALEPH
Collaborations \cite{CLEO,ALEPH} (see Fig.~\ref{fig:data}). To obtain
${\rm d}R_\tau/{\rm d}s$, we multiply the normalized distributions by
the world average $R_\tau=R_\tau(m_\tau^2)=3.642\pm 0.010$
\cite{PDG95}. Not shown is the contribution from $\tau\to
h^-\nu_\tau$ with $h^-=\pi^-$ or $K^-$, which has a branching ratio
of $(11.77\pm 0.14)\%$ \cite{Callot}. We integrate these spectra over
$s$, combine the results weighted by their statistical errors, and
add the systematic errors, which we estimate by taking the difference
between the CLEO and ALEPH data.\footnote{As the ALEPH data are
preliminary, this estimate may be taken with caution. However, since
inclusive quantities such as $R_\tau(s_0)$ do only probe gross
features of the hadronic mass distribution, systematic errors play a
minor role in our analysis.}
This is justified, since the dominant sources of systematic errors
are different in the two analyses. The result is shown in
Fig.~\ref{fig:data}. It is represented as a band, since the errors in
the $R_\tau(s_0)$ values are strongly correlated. The two curves show
theoretical calculations of $R_\tau(s_0)$ based on the OPE approach
outlined above. The solid line is obtained using fixed-order
perturbation theory to order $\alpha_s^4$. The dashed line is
obtained by adding to this a resummation of renormalon chains of
order $\alpha_s^5$ and higher, using the results of Ref.~\cite{MN}.
The value of $\alpha_s(m_\tau^2)$ has been adjusted so as to fit the
data at $s_0=m_\tau^2$. The central values obtained in the two
schemes are $\alpha_s(m_\tau^2)=0.329$ (fixed-order) and
$\alpha_s(m_\tau^2)=0.309$ (resummed). Their difference provides an
estimate of the uncertainty due to unknown higher-order corrections,
which is more conservative than that obtained by omitting the term of
order $\alpha_s^4$ in the fixed-order calculation. Varying the values
of the nonperturbative parameters within conservative limits changes
$\alpha_s(m_\tau^2)$ by up to 2\%. Adding linearly the perturbative
uncertainty ($\pm 0.020$), the nonperturbative uncertainty ($\pm
0.006$), and the experimental uncertainty ($\pm 0.004$), we find
\begin{eqnarray}
   \alpha_s(m_\tau^2) &=& 0.329\pm 0.030 \,, \nonumber\\
   \alpha_s(m_Z^2) &=& 0.119\pm 0.004 \,.
\end{eqnarray}
For the sake of completeness, we have translated our result into a
value of $\alpha_s$ at the mass of the $Z$ boson.

The assumption of global quark--hadron duality can be tested by
comparing the data for the quantity $R_\tau(s_0)$ at values
$s_0<m_\tau^2$ with the theoretical prediction \cite{LP2}. Given
$\alpha_s(m_\tau^2)$, the value of $\alpha_s(s_0)$ follows from the
solution of the renormalization-group equation
\begin{eqnarray}\label{RGE}
   \mu^2\,{{\rm d}\alpha_s(\mu^2)\over{\rm d}\mu^2}
   &=& -\alpha_s(\mu^2)\,\beta[\alpha_s(\mu^2)] \,, \nonumber\\
   \beta(\alpha_s) &=& \beta_0\,{\alpha_s\over 4\pi}
    + \beta_1\,\bigg( {\alpha_s\over 4\pi} \bigg)^2
    + \beta_2\,\bigg( {\alpha_s\over 4\pi} \bigg)^3 + \dots \,,
\end{eqnarray}
where $\beta_0=9$, $\beta_1=64$ and $\beta_2=3863/6$ are the first
three coefficients of the $\beta$-function, evaluated for $n_f=3$
light quark flavours. (The value of $\beta_2$ is specific to the
$\overline{\mbox{\sc ms}}$ scheme.) Whereas the theoretical
uncertainties are a limiting factor in the determination of
$\alpha_s(m_\tau^2)$, they have little influence on the $s_0$
dependence of $R_\tau(s_0)$. For the perturbative part of the
calculation this is apparent from the good agreement of the two
curves in Fig.~\ref{fig:data}, which refer to values of
$\alpha_s(m_\tau^2)$ that differ by 9\%. Varying the values of the
nonperturbative parameters has a negligible effect ($\sim 0.5\%$ at
$s_0=1$~GeV$^2$) on the value of $R_\tau(s_0)$ for $s_0<m_\tau^2$,
since the only dependence on $s_0$ in (\ref{Rpower}) comes from the
quark-mass corrections, which are known with higher accuracy than the
vacuum condensates. Hence, the $s_0$ dependence of $R_\tau(s_0)$ is
predicted essentially without free parameters, and the comparison
between theory and experiment provides a direct test of quark--hadron
duality. We find good agreement over the entire range
0.7~GeV$^2<s_0<m_\tau^2$, indicating that in $\tau$ decays duality
holds as soon as the integral over the hadronic mass distribution
includes the $\rho$ resonance peak. The onset of duality happens
almost instantaneously, in accordance with general expectations.

\begin{figure}
   \epsfysize=9cm
   \centerline{\epsffile{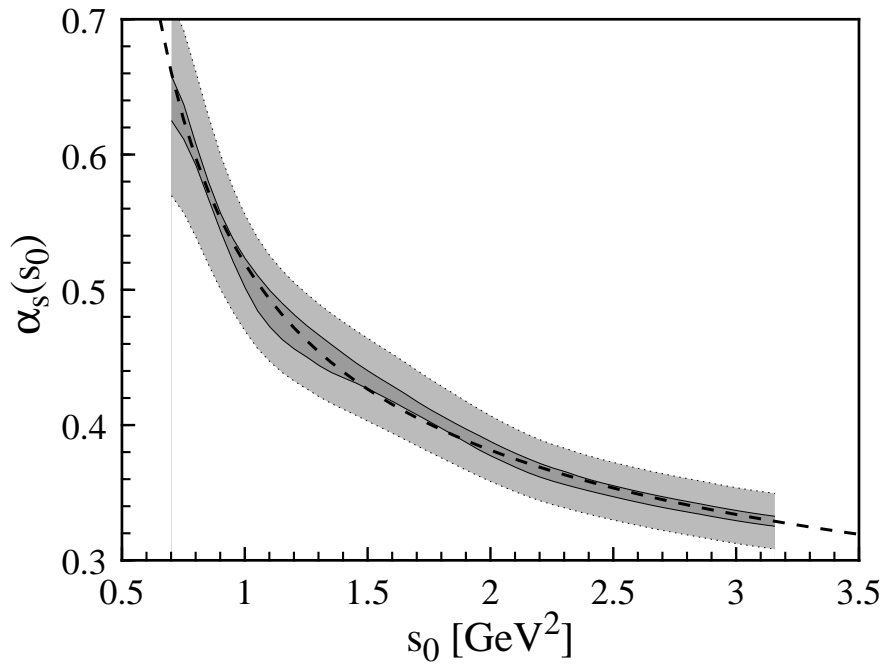}}
   \centerline{\parbox{14cm}{\caption{\label{fig:alphas}
Values of $\alpha_s(s_0)$ extracted from the data on $R_\tau(s_0)$.
The dark band represents the experimental errors, the light one the
sum of the experimental and theoretical errors. The errors are
strongly correlated. The dashed line shows the three-loop QCD
prediction for the running coupling constant.
}}}
\end{figure}

From now on, we shall rely on this behaviour and assume that for
$s_0>0.7$~GeV$^2$ possible violations of duality can be neglected. We
then turn to the main focus of our study: a test, at low energies, of
the QCD prediction (\ref{RGE}) for the running of the coupling
constant. $\tau$ decays are an ideal place to study this phenomenon
since the value of $\alpha_s(s_0)$ changes by a factor~2 over the
region where duality holds, which is about the same change as in the
region between 5 and 100~GeV. From the quantity $R_\tau(s_0)$ shown
in Fig.~\ref{fig:data}, we extract $\alpha_s(s_0)$ as a function of
$s_0$ by fitting to the data the theoretical prediction obtained from
(\ref{Rs0}), (\ref{Mkpert})--(\ref{Rpower}). The result, including
experimental errors only, is represented by the dark band in
Fig.~\ref{fig:alphas}. Theoretical uncertainties arise from the
truncation of the perturbation series and from the uncertainty in the
values of the nonperturbative parameters. As discussed above, they
affect the overall scale of the $\alpha_s$ values (by about 8--10\%),
but have very little effect on the evolution of the coupling
constant. The sum of the experimental and theoretical errors is
represented by the light band. The dashed curve shows the QCD
predictions for $\alpha_s(s_0)$ obtained at three-loop order,
normalized to the central value of the data at $s_0=m_\tau^2$. The
observed scale dependence of the running coupling constant is in good
agreement with the QCD prediction. The small oscillation of the
experimental band around the theoretical curve, which could be due to
some deviations from duality in the $a_1$ region, are not significant
given the precision of the data.

\begin{figure}
   \epsfysize=9cm
   \centerline{\epsffile{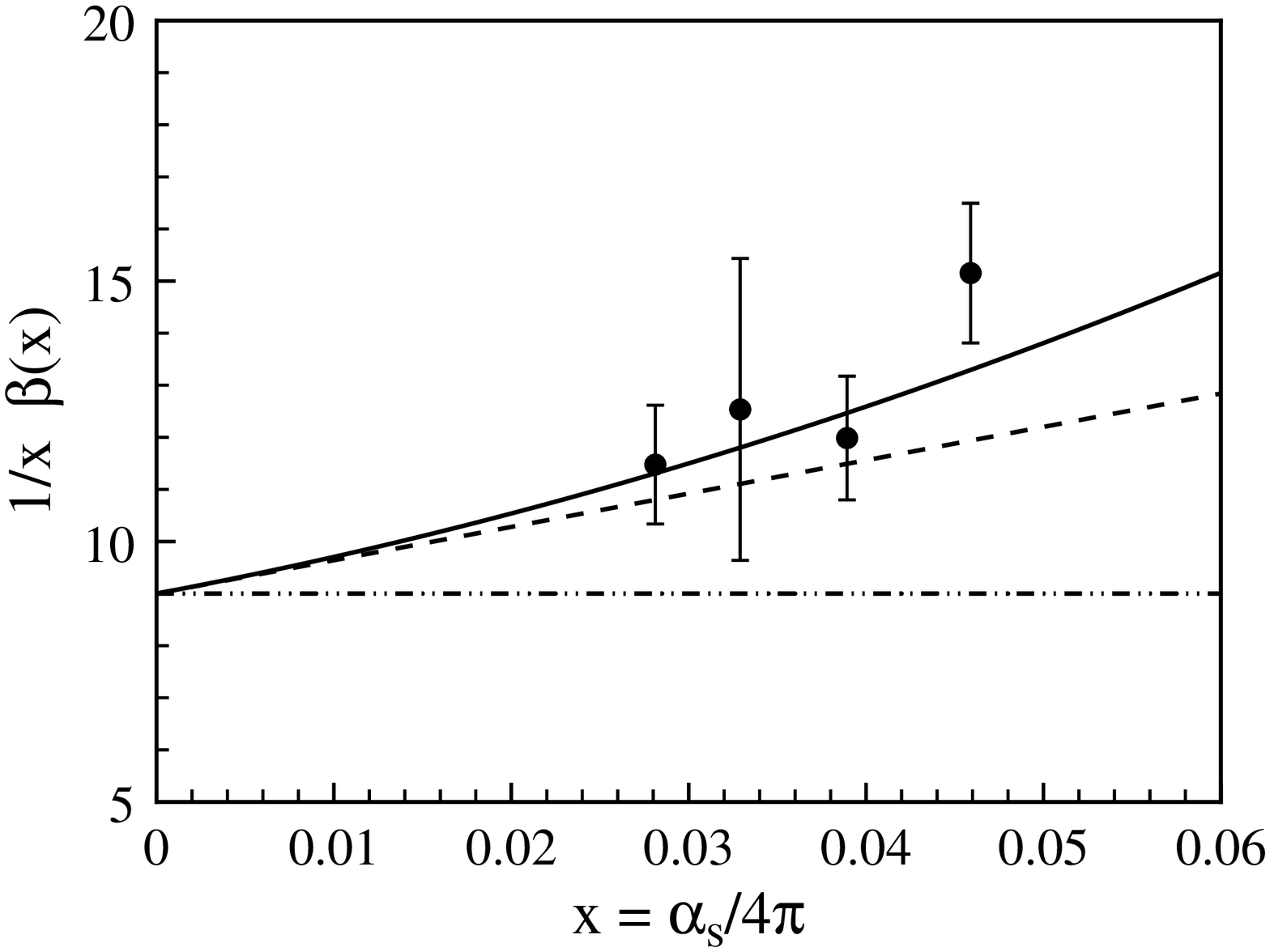}}
   \centerline{\parbox{14cm}{\caption{\label{fig:betafun}
Experimental determination of the $\beta$-function. The curves show
the QCD prediction at one-loop (dash-dotted), two-loop (dashed) and
three-loop (solid) order.
}}}
\end{figure}

To quantify this agreement, we extract from the data the
$\beta$-function that describes according to (\ref{RGE}) the running
of $\alpha_s(s_0)$. Defining $x=\alpha_s(s_0)/4\pi$, we have
\begin{equation}\label{beta}
   - {4\pi\over\alpha_s^2(s_0)}\,
   {{\rm d}\alpha_s(s_0)\over{\rm d}\ln s_0} = {\beta(x)\over x}
   = \beta_0 + \beta_1 x + \beta_2 x^2 + \dots \,.
\end{equation}
We approximate the derivative ${\rm d}\alpha_s/{\rm d}\ln s_0$ by a
ratio of differences, $\Delta\alpha_s/\Delta\ln s_0$, for a set of
$s_0$ values chosen such that the differences $\Delta\alpha_s$ are
large enough to be significant given the errors in the measurement.
For $\alpha_s(s_0)$ in (\ref{beta}) we take the central value of each
interval. We use the following $s_0$ values: 0.75, 0.95, 1.35, 2.06,
and 3.16~GeV$^2$, corresponding to four intervals of increasing width
$\Delta\ln s_0$, but constant $\Delta\alpha_s\simeq 0.075$. The
results are shown in Fig.~\ref{fig:betafun}. The estimate of the
errors includes the theoretical uncertainties, the error due to the
choice of finite intervals in $\alpha_s$, and the experimental
errors, which in this case are the dominant ones. The curves in
Fig.~\ref{fig:betafun} show the QCD $\beta$-function at one-, two-
and three-loop order in perturbation theory. The data provide clear
evidence for the running of the coupling constant. Moreover, they
prefer a running that is stronger than predicted at one-loop order.
Between the three curves, the one that shows the three-loop
prediction provides the best description of the data. Performing a
fit with the three-loop $\beta$-function, where $\beta_0=9$ and
$\beta_1=64$ are kept fixed but the three-loop coefficient $\beta_2$
is treated as a parameter, we find $\beta_2^{\rm exp}/\beta_2^{\rm
th}=1.6\pm 0.7$. We believe that such an experimental determination
of the $\beta$-function beyond the leading order can at present be
done only in $\tau$ decays. (A high-precision measurement of $R_{e^+
e^-}(s)$ in the region below the charmonium resonances would provide
an alternative place for such a study.) At higher energies, the value
of $\alpha_s$ is too small to distinguish between the three curves in
Fig.~\ref{fig:betafun}; measurements in the region $Q\sim 100$~GeV,
for instance, correspond to values $x\sim 0.01$.

In summary, we have presented a method to measure the running
coupling constant $\alpha_s(s)$ in the low-energy region
$0.7~\mbox{GeV}^2<s<m_\tau^2$ using $\tau$ decay data obtained in a
single experiment. It provides a test of one of the key features of
QCD in a region where the effect of the running of $\alpha_s$ is most
pronounced. The theoretical analysis is based on the OPE and the
assumption of global quark--hadron duality. We have tested this
assumption and find that it seems to hold if the $\tau$ decay rate is
integrated over an energy interval large enough to include the $\rho$
resonance peak. Our analysis provides a test of QCD at scales
comparable with the lowest ones attainable before ($Q^2\simeq
2.5$~GeV$^2$ in deep-inelastic scattering), and with higher precision
than all other single measurements of the running to date. We have
extracted for the first time the $\beta$-function from data and find
that it is in good agreement with the three-loop prediction of QCD.

\vspace{0.3cm}
{\it Acknowledgements:\/}
We would like to thank R.K.~Ellis, M.~Mangano, P.~Nason,
C.T.~Sach\-rajda and R.~Sommer for useful discussions.

\newpage

\newpage
\begin{thebibliography}{99}

\bibitem {Gros}
D.J. Gross and F. Wilczek, Phys.\ Rev.\ Lett.\ {\bf 30}, 1343
(1973);
H.D. Politzer, {\it ibid.\/} {\bf 30}, 1346 (1973).

\bibitem {Webb}
B. Webber, in: Proc.\ 27th Int.\ Conf.\ on High Energy Physics,
Glasgow, Scotland, July 1994, eds.\ P.J.~Bussey and I.G.~Knowles (IOP
Publ., Bristol, 1995), Vol.~1, p.~213.

\bibitem {Giel}
M. Derrick {\it et al.\/} (ZEUS Coll.), Phys.\ Lett.\ B {\bf 363},
201 (1995);
W.T. Giele, E.W.N. Glover and J. Yu, preprint FERMILAB-Pub-95/127-T
(1995) [hep-ph/9506442].

\bibitem {SVZ}
M.A. Shifman, A.I. Vainsthein and V.I. Zakharov, Nucl.\ Phys.\ B
{\bf 147}, 385 and 448 (1979).

\bibitem {Rtau1}
E. Braaten, S. Narison and A. Pich, Nucl.\ Phys.\ B {\bf 373}, 581
(1992).

\bibitem {PQW}
E.C. Poggio, H.R. Quinn and S. Weinberg, Phys.\ Rev.\ D {\bf 13},
1958 (1976).

\bibitem {epem}
S.I. Eidelman, L.M. Kurdadze, and A.I. Vainshtein, Phys.\ Lett.\ B
{\bf 82}, 278 (1979).

\bibitem {LP2}
F. Le Diberder and A. Pich, Phys.\ Lett.\ B {\bf 289}, 165 (1992).

\bibitem {MN}
M. Neubert, Nucl.\ Phys.\ B {\bf 463}, 511 (1996).

\bibitem {Sirl}
W.J. Marciano and A. Sirlin, Phys.\ Rev.\ Lett.\ {\bf 56}, 22 (1986).

\bibitem {ECH}
P.M. Stevenson, Phys.\ Rev.\ D {\bf 23}, 2916 (1981);
G. Grunberg, Phys.\ Lett.\ B {\bf 221} 70 (1980); Phys.\ Rev.\ D {\bf
29}, 2315 (1984).

\bibitem {Kata}
A.L. Kataev and V.V. Starshenko, Mod.\ Phys.\ Lett.\ A {\bf 10}, 235
(1995).

\bibitem {tHof}
G. 't Hooft, in: Proc.\ 15th Int.\ School of Subnuclear Physics,
Erice, Sicily, 1977, ed.\ A.~Zichichi (Plenum Press, New York, 1979),
p.~943.

\bibitem {BBB}
P. Ball, M. Beneke and V.M. Braun, Nucl.\ Phys.\ B {\bf 452}, 563
(1995).

\bibitem {CLEO}
T. Coan {\it et al.\/} (CLEO Coll.), Phys.\ Lett.\ B {\bf
356}, 580 (1995).

\bibitem {ALEPH}
L. Duflot, Nucl.\ Phys.\ B (Proc.\ Suppl.) {\bf 40}, 37 (1995).

\bibitem {PDG95}
L. Montanet {\it et al\/.} (Review of Particle Properties), Phys.\
Rev.\ D {\bf 50}, 1173 (1994), and 1995 off-year partial update (URL:
{\tt http://pdg.lbl.gov/}).

\bibitem {Callot}
R. Stroynowski, Nucl.\ Phys.\ B (Proc.\ Suppl.) {\bf 40}, 569 (1995).

\end{thebibliography}
\end{document}